# Approaching Irrational Quantization

Zhihong Wu[1], Zhimou Chen[1], Siyuan Zhang[1], Jinyang Chen[1], L. N. Pfeiffer[2], K. W. West[2], Yang Liu[1,3], and Xi Lin[1,3,4,*]

[1]International Center for Quantum Materials, School of Physics, Peking University, Beijing, 100871, China

[2]Department of Electrical Engineering, Princeton University, Princeton, New Jersey, 08544, USA

[3]Hefei National Laboratory, Hefei, 230088, China

[4]Interdisciplinary Institute of Light-Element Quantum Materials and Research Center for Light-Element Advanced Materials, Peking University, Beijing, 100871, China

* xilin@pku.edu.cn

*The integer and fractional quantum Hall effects reveal that quantized physical responses can take integer and fractional values, respectively. This naturally raises the question of whether quantization can extend beyond rational numbers to an irrational value. Here we show that such a quantized response can emerge from the limiting behavior of a finite, experimentally realizable geometric network. We develop a method based on infinite geometric structures, in which a sequence of finite networks rapidly converges toward an infinite limit. Following this approach, we identify a planar network whose normalized resistance converges to the irrational value* $\sqrt{3}$*, and realize a finite approximation with high-precision geometric control. The measured normalized resistance is* $1.735 \pm 0.001\ h/e^2$ *for a fully occupied Landau level in the designed pattern. Our experimental results extend the concept of quantization beyond rational values and demonstrate that an irrational quantized response can be approached through a finite physical system.*

Quantization of physical observables is a hallmark of quantum phenomena, with particularly striking manifestations in quantum transport. The integer quantum Hall (IQH) effect in high-quality two-dimensional electron systems provides a robust quantized Hall conductance, $G_{xy} = Ne^2/h$, where $N$ is an integer[1-3]. Its robustness against disorder and sample geometry, together with the underlying chiral edge states and topology, has established quantum Hall systems as a paradigmatic platform for quantized transport and enabled precision resistance metrology through the von Klitzing constant $R_{\mathrm{K}}$ $(h/e^2)$. The fractional quantum Hall (FQH) effect further extends quantized Hall conductance to fractional filling factors[4], revealing fractionalized quasiparticles and new forms of strongly correlated many-body physics, including the possibility of non-Abelian statistics in certain states[5-8]. Quantum Hall physics has thus established quantized resistance at integer and, through many-body effects, at fractional and more generally, rational values. Yet rational and irrational numbers together exhaust the real numbers, leaving open whether quantization can extend into the irrational sector. This progression naturally raises a fundamental question: can a quantum resistance plateau be characterized by an irrational number?

The integer quantization of the IQH effect can be understood from both transport and topological perspectives[2-3]. In the Landauer-Büttiker formalism[9-10], each perfectly transmitting chiral edge channel contributes $e^2/h$ to the conductance, leading naturally to an integer number of channels. From the bulk perspective, the Hall conductance is related to the Chern number, an integer-valued topological invariant. The FQH effect extends quantum Hall quantization beyond this single-particle integer picture through many-body correlations[11-14], with the composite-fermion picture providing an intuitive understanding of a broad class of fractional states and hence rational quantized values[15-17]. Thus, the established mechanisms of quantum Hall quantization are intrinsically connected to integer-valued topology or many-body mechanisms leading to rational values, but provide no obvious route toward a quantized response characterized by an irrational value. A fundamentally different mechanism is therefore required to extend quantization beyond the rational numbers.

Several quantum systems have demonstrated how non-periodic and complex geometric structures can generate unconventional physical properties. Quasiperiodic and quasicrystalline systems naturally involve irrational ratios and aperiodic long-range order[18], while Hofstadter-type systems exhibit rich spectral structures associated with irrational flux values[19]. However, the presence of irrational structural parameters does not by itself imply a quantized response with an irrational value. In physical systems, observable properties are often determined not only by individual elements but also by their collective organization and geometric arrangement. Whether the geometric organization of quantized elements can produce a quantized response with an irrational value remains an open question.

In this letter, we develop a method based on infinite geometric structures, in which finite realizations converge toward an irrational limiting value of the quantized response. Using this approach, we identify a finite planar structure that is experimentally accessible and realize it with quantum Hall resistance elements. The measured quantized resistance is consistent with the predicted limiting value with experimental accuracy.

We consider an infinite two-dimensional network with a periodic structure, as schematically illustrated in Fig. 1. A crystal lattice provides a natural example of such a structure, in which nodes are arranged periodically in space and interact with other nodes within a finite range. More generally, the nodes may represent physical degrees of freedom located at different positions in a two-dimensional periodic structure, with local couplings connecting neighboring or nearby nodes. When these couplings are linear, the response at each node can be expressed as a linear combination of the inputs at other nodes. The complete set of coupling relations can therefore be represented by matrices.

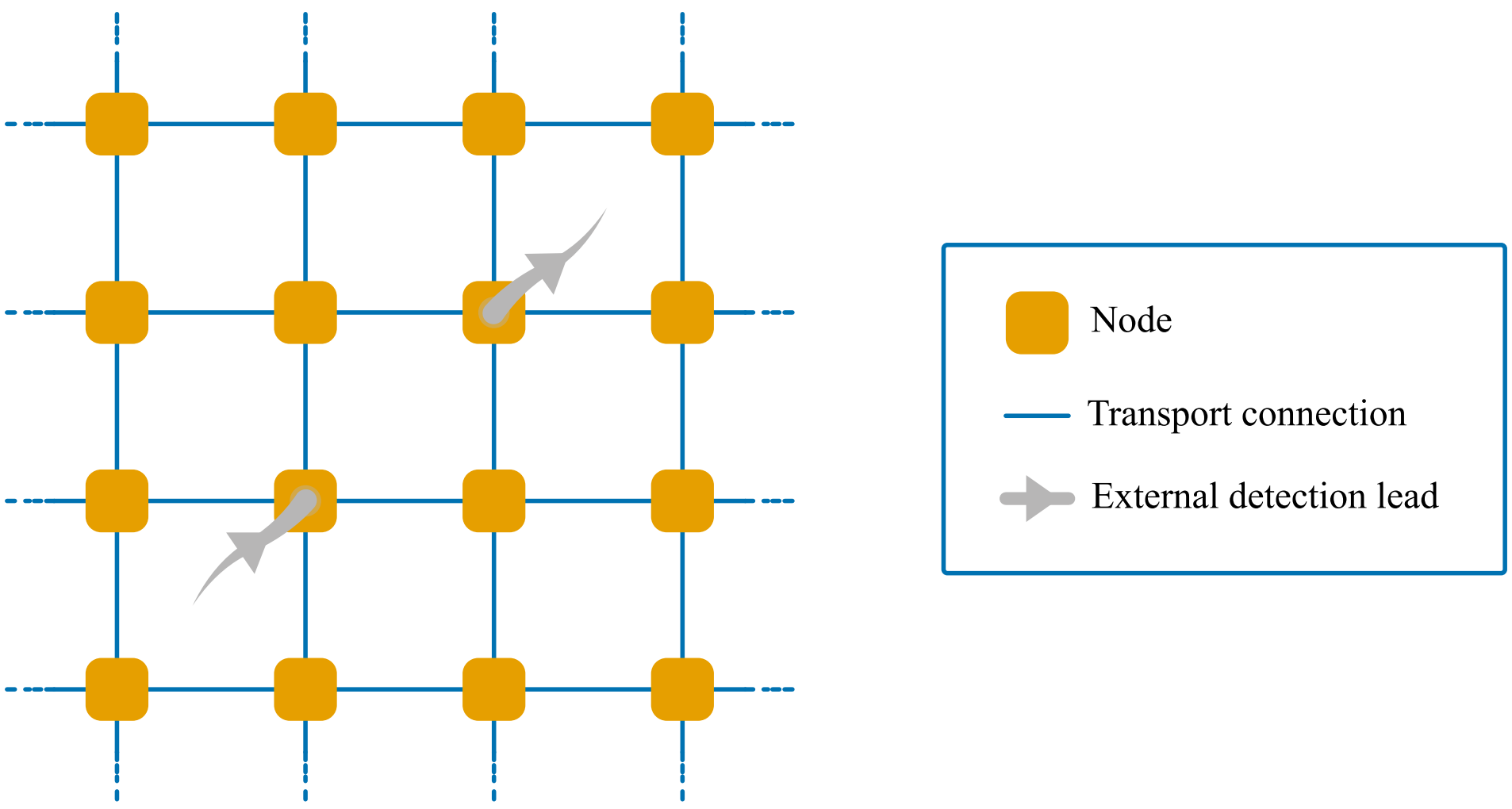

**Figure 1: A schematic example of a two-dimensional network.** A typical square lattice is shown as a general model, with only one type of node in the primitive cell. For the infinite square lattice, if all connections have the same resistance $R_0$, the equivalent resistance between two nodes can be written as $(a + b/\pi)R_0$, where $a$ and $b$ are rational numbers.

The periodicity of the network allows its response to be decomposed into independent spatial Fourier modes[20], each characterized by a wave vector $\boldsymbol{k}$. For an infinite periodic system, these modes form a continuum within the first Brillouin zone (BZ), although the response matrix may become singular at a few isolated points. In the Fourier representation, the linear response of the network is described by a matrix $Y(\boldsymbol{k},\omega)$, such that the response to an applied source satisfies $I(\boldsymbol{k},\omega) = Y(\boldsymbol{k},\omega)V(\boldsymbol{k},\omega)$. For example, in an AC circuit, the linear response can be represented by an admittance matrix $Y(\boldsymbol{k},\omega)$, relating the currents to the node voltages. Where $Y(\boldsymbol{k},\omega)$ is nonsingular, its inverse matrix therefore determines the response of the network to the applied source. For a source and measurement configuration described by $\boldsymbol{p}(\boldsymbol{k})$, the corresponding response can be written as Eq. 1 (see Supplementary Material, Section A).

$$F_p(\omega) = \int_{\mathrm{BZ}} \frac{\mathrm{d}^d\boldsymbol{k}}{(2\pi)^d} \boldsymbol{p}^\dagger(\boldsymbol{k}) Y^{-1}(\boldsymbol{k},\omega)\boldsymbol{p}(\boldsymbol{k}) \tag{1}$$

Here, $\boldsymbol{p}(\boldsymbol{k})$ specifies the source and measurement configuration, while $Y(\boldsymbol{k},\omega)$ describes the linear couplings within and between primitive cells. The dimensionality $d$ determines the dimensionality of the BZ. The choice of primitive cell determines the internal degrees of freedom of the network. The number of nodes within the primitive cell sets the order of $Y(\boldsymbol{k},\omega)$, while the couplings within and between neighboring cells determine its detailed $\boldsymbol{k}$-dependence. Increasing the number of nodes and couplings within the primitive cell therefore provides additional degrees of freedom for engineering the functional form of the response.

A natural physical realization of well-defined resistive elements is provided by ballistic transport systems, where transport can become quantized under appropriate conditions. In a ballistic conductor, the conductance is described by the Landauer formula[9-10], as shown in Eq. 2.

$$G = \frac{e^2}{h}\sum_n T_n \tag{2}$$

Here $T_n$ is the transmission probability of the $n$-th channel. When the relevant channels are fully transmitted, $T_n = 1$, the conductance is quantized in integer multiples of $e^2/h$. Quantum Hall systems provide particularly robust realizations for quantized transport, with IQH states and FQH states giving precisely quantized Hall conductance. The corresponding quantized resistances provide experimentally accessible circuit elements with precisely defined rational values when expressed in appropriate resistance units[1,4]. This makes quantized transport a natural physical platform for constructing periodic networks whose local circuit parameters are rational.

The mathematical structure of $F_p(\omega)$ is not determined solely by the numerical values of the building blocks. When the circuit parameters are rational, the coefficients in $Y(\boldsymbol{k},\omega)$ are rational, while its $\boldsymbol{k}$-dependence is determined by the periodic geometry of the network. For fixed $\boldsymbol{k}$ and $\omega$, matrix inversion preserves the rationality of these coefficients wherever $Y(\boldsymbol{k},\omega)$ is nonsingular, while the resulting response remains a function of $\boldsymbol{k}$ and $\omega$. The subsequent integration over the BZ is fundamentally different, as it maps these $\boldsymbol{k}$-dependent response functions to the physical observable $F_p(\omega)$, and the resulting value need not be rational even when all local circuit parameters are rational. The mathematical structure of $F_p(\omega)$ is therefore determined not only by the local building blocks and their couplings, but also by the dimensionality and geometry of the infinite periodic network.

We now turn to a specific physical realization of this general framework. We consider the purely resistive limit, for which $\omega = 0$ and $F_p$ reduces to an effective resistance. In this limit, the response is determined solely by the resistive elements and their network connectivity, providing a direct physical realization of the mathematical structure described above. For a finite realization of the network, however, the deviation from the infinite-size limit becomes a crucial issue. We therefore approximate the infinite periodic structure by a finite network containing a large but finite number of primitive cells. Let $R_L$ denote the effective resistance of a finite-size network with characteristic linear size $L$, for example, the side length of a hypercube, and let $R_\infty$ denote the corresponding infinite-system value under the limiting procedure. The finite-size error is given by $E_L = R_L - R_\infty$ (see Supplementary Material, Section B). For a sequence of finite networks approaching the infinite system with $L$ increases, $R_L$ approaches $R_\infty$, and the rate of this convergence determines how efficiently a finite network can reproduce the response of the infinite system. The convergence of a finite resistor network toward its infinite-size limit depends on how the network geometry distributes the current throughout the structure. In a finite network, the boundaries inevitably modify the current distribution, and their influence decreases as the network size increases. The rate at which this boundary effect diminishes is therefore determined by the connectivity and spatial structure of the network. Different network geometries can consequently exhibit markedly different finite-size convergence, even when they are constructed from the same resistive elements.

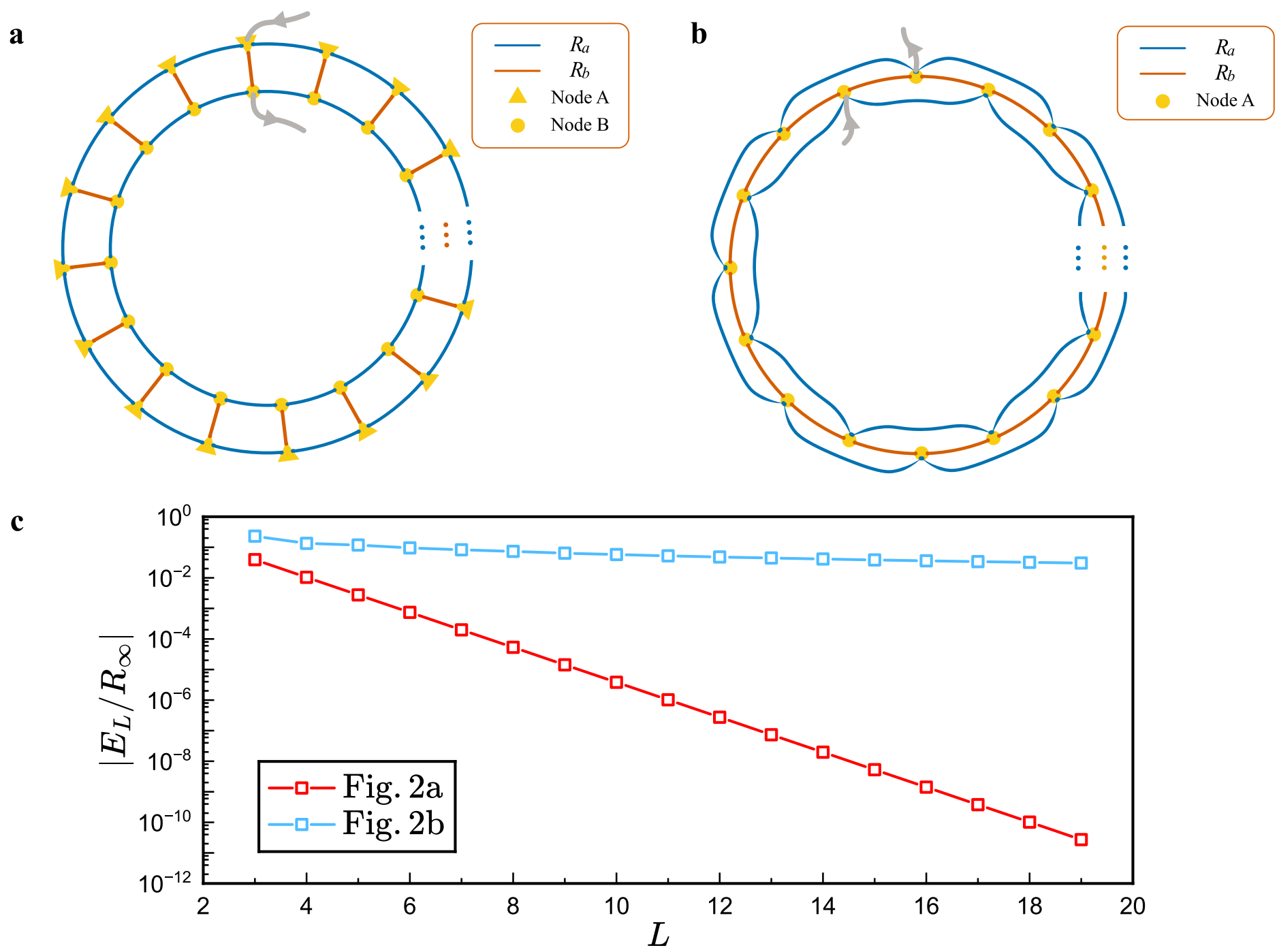

**Figure 2: Examples for different forms of the leading error term of the network. a.** In a two-layer concentric ring exhibiting exponential decay with only nearest-neighbor interactions, as $e^{-\kappa L}$. There are two distinguishable types of nodes labeled as A and B. **b.** In a ring exhibiting power-law decay with both nearest-neighbor and next-nearest-neighbor interactions, as $1/L$. Compared to Fig. 2a, only one type of node exists. **c.** Comparison of decay rates between cases in Fig. 2a and Fig. 2b. Here we take $R_a = R_b = 3R_0$ in Fig. 2a and $R_a = 6R_0, R_b = 3R_0$ in Fig. 2b. Their limiting resistances $R_\infty$ are both $\sqrt{3}R_0$.

To investigate finite-size convergence in a simpler setting, we consider a one-dimensional reduction of the general network introduced in Fig. 1. Within this simplified structure, we consider two possible configurations, referred to as the exponential-decay configuration and the power-law-decay configuration, respectively, as shown in Fig. 2a and Fig. 2b. Although the two configurations are constructed from the same resistive elements, their different connection arrangements lead to distinct finite-size convergence behaviors. We calculate the effective resistance for finite networks of increasing size and compare the results with their infinite-size limits. For both configurations, the finite-size convergence is determined by the analytic structure of the corresponding network response. The different connection configurations therefore provide a means of controlling the analytic structure of the response and, consequently, the convergence rate.

For the one-dimensional configuration shown in Fig. 2b, the resistance is measured between two adjacent nodes, exhibiting power-law convergence. Each node can be regarded as a primitive cell of a one-dimensional crystal, so $L$ represents the number of primitive cells in the finite-sized network. Since the overall potential shift at $\boldsymbol{k} = 0$ generally leads to the non-invertibility of $Y(\boldsymbol{k})$, the asymptotic expansion near $\boldsymbol{k} = 0$ determines the form of the finite-size error. When $R_a = 6R_0, R_b = 3R_0$, the general term can be written as $R_L = \sqrt{3}R_0\left[1 + \frac{2}{(-2-\sqrt{3})^L - 1}\right] - \frac{R_0}{L}$ (see Supplementary Material, Section B). Therefore, the asymptotic form of the error is $E_L \sim -\frac{R_0}{L}$.

Although the configuration shown in Fig. 2b exhibits power-law convergence, a different measurement configuration can yield much faster convergence, as illustrated in Fig. 2c. For the one-dimensional exponential-decay configuration shown in Fig. 2a, the measurement is conducted between two adjacent nodes on the inner and outer rings. These two nodes can be collectively regarded as a primitive cell. Specifically, under the measurement configuration shown in Fig. 2a, the specific structural design and node selection combine to suppress the contribution associated with the $\boldsymbol{k} = 0$ mode, thereby removing the source of leading finite-size error. When $R_a = R_b = 3R_0$, the general term can be written as $R_L = \sqrt{3}R_0\left[1 + \frac{2}{(2+\sqrt{3})^L - 1}\right]$ (see Supplementary Material, Section B). Therefore, the leading finite-size error is exponentially small, $E_L \sim 2\sqrt{3}R_0\left(2+\sqrt{3}\right)^{-L}$. Under this measurement, the rate of decay is relatively rapid.

Among various networks, exponential and power-law forms represent two types of decay with vastly different convergence rates. This distinction arises from the interplay between network structure and the nodes being measured. Power-law decay is generally expected in general networks, particularly under periodic boundary conditions where $1/L^d$ multiplied by an additional constant factor serves as the primary decay component, showing that for finite networks, the error generally relates directly to the overall size of the finite-size network. For example, if we measure between two adjacent nodes on the same outer ring in Fig. 2a, then the general form of $R_L$ will become $\left[3-\frac{\sqrt{3}}{2}-\frac{3}{2L}-\frac{\sqrt{3}}{(2+\sqrt{3})^L-1}\right]R_0$ (see Supplementary Material, Section B). Power-law decay follows the $1/L$ leading term in this expression, which is associated with the singular behavior of the integrand in Eq. 1 is singular at $\boldsymbol{k}=0$.

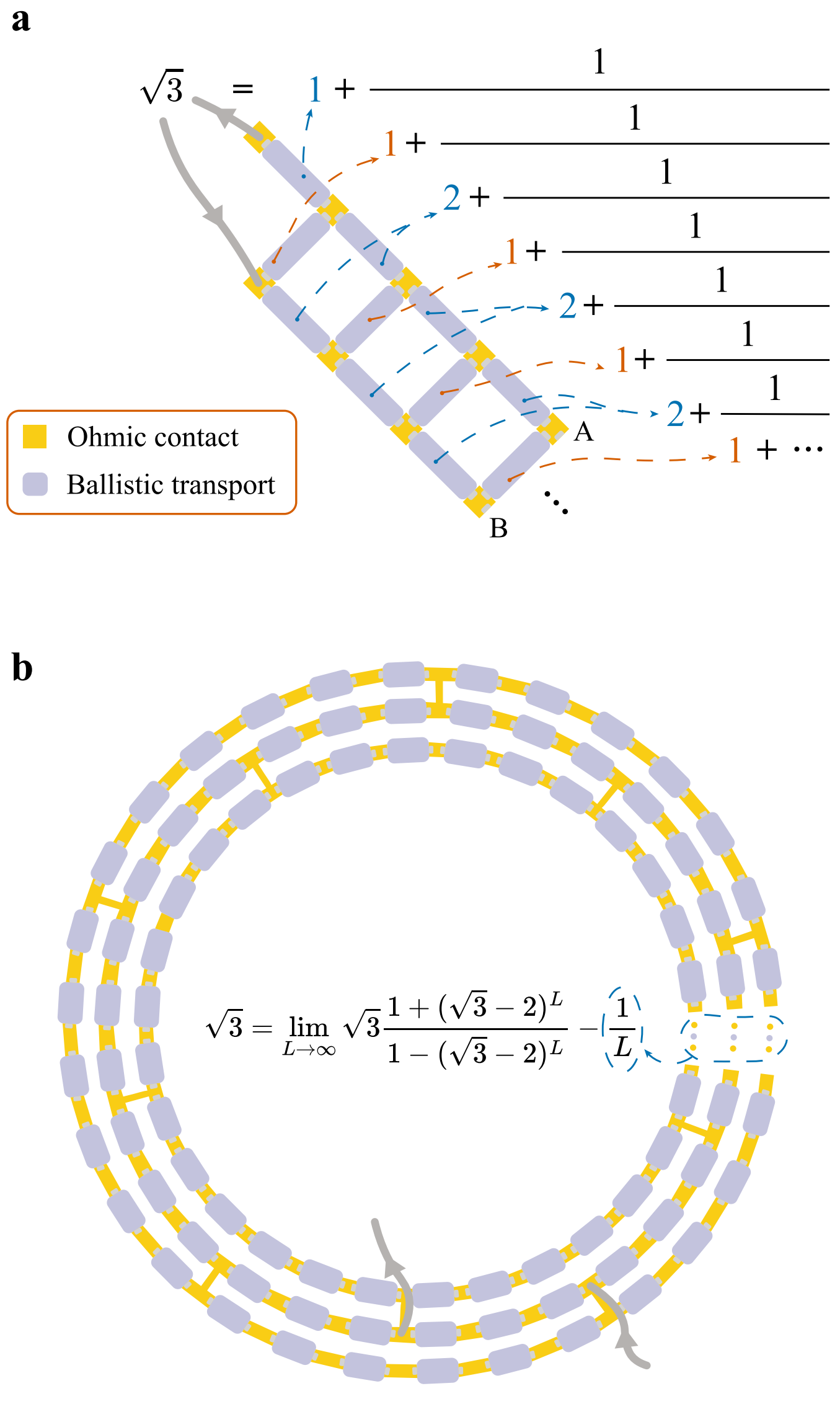

**Figure 3: Network design diagram of multiple components of the same type. a.** A semi-infinite one-dimensional ladder structure with an exponential decay (see Supplementary Material, Section B). This can be viewed as Fig. 2a cut in half. **b.** A three-layer concentric ring structure with a power-law decay. Nearest-neighbor connection is $3R_0$, and next-nearest-neighbor connection is $6R_0$.

To experimentally realize the irrationally quantized network, we use the quantized resistance plateaus of the IQH effect and FQH effect in a two-dimensional electron gas (2DEG) as the basic building blocks. As illustrated in Fig. 3a, we implement a finite network with $L=4$, constrained by the sample size and the practical difficulty of fabricating a large number of ohmic contacts. The calculated resistance of this finite network differs from the infinite-network limit by only approximately $4 \times 10^{-6}$ in relative terms (see Supplementary Material, Table S1), reflecting the rapid finite-size convergence of the designed structure. This deviation is well below the experimental uncertainty in this work, indicating that the $4 \times 10^{-6}$ network provides a satisfactory realization of the theoretically defined infinite structure. Moreover, the structure of Fig. 3b is noticeably more difficult to realize.

We characterize a conventional six-terminal Hall bar fabricated on the same sample as a control. At 300 mK, several well-developed IQH plateaus are observed, as shown in Fig. 4b. This establishes the accuracy of the measurement system and provides a reference for the network measurements under the same conditions. We then measure the quantum network. For a fully occupied Landau level with filing factor one ($\upsilon = 1$), the network resistance is $1.735 \pm 0.001\ h/e^2$, compared with the predicted infinite-network value $\sqrt{3}\ h/e^2$. The measured value is slightly higher than the theoretical value, as expected because the finite ohmic contact resistance contributes positively to the measured value. Since the finite-size correction of the $L = 4$ network is only approximately $4 \times 10^{-6}$, much smaller than the experimental uncertainty, the observed deviation from the ideal value is dominated by experimental factors rather than by the finite size of the network.

We further examine the stability of the network quantization by varying the temperature. As shown in Fig. 4a, well-developed quantized resistance plateaus are observed as a function of magnetic field at $\upsilon = 2$. Measurements performed at 300, 600, and 1200 mK show that the quantized network resistance remains essentially unchanged over this temperature range. This temperature stability indicates that the observed quantization is robust against thermal variation within the investigated range.

A more direct and systematic test of the predicted irrational quantization is obtained by examining the scaling of the resistance with the inverse filling factor. As shown in Fig. 4b, the conventional integer Hall resistance follows the expected linear dependence on $1/\upsilon$, while the network resistance for the IQH states follows the same dependence with a slope smaller by a factor of $\sqrt{3}$. The network data at $\upsilon = 1, 2, 3, 4$ consistently follow this scaling relation, demonstrating that the irrational coefficient is a systematic consequence of the network structure rather than an isolated feature of a particular quantum Hall plateau. The agreement across multiple integer filling factors provides a quantitative verification of the predicted irrational scaling.

We also investigate the network using the FQH states at $\upsilon = 4/3,\ 5/3$. The corresponding Hall resistances remain consistent with the same $1/\upsilon$ scaling established by the IQH states, indicating that the fractional states themselves are quantized with acceptable experimental uncertainty. In contrast, the network resistances at these fractional filling factors deviate from the $\sqrt{3}$ scaling established by the integer states beyond the experimental uncertainty (see Supplementary Material, Section C and Fig. S1). At temperatures around 60 mK, the finite contact resistance and reduced contact quality may have a stronger influence on FQH transport and edge-state equilibration. Alternatively, the deviation may reflect additional transport mechanisms associated with FQH states that are not captured by treating them as ideal circuit elements.

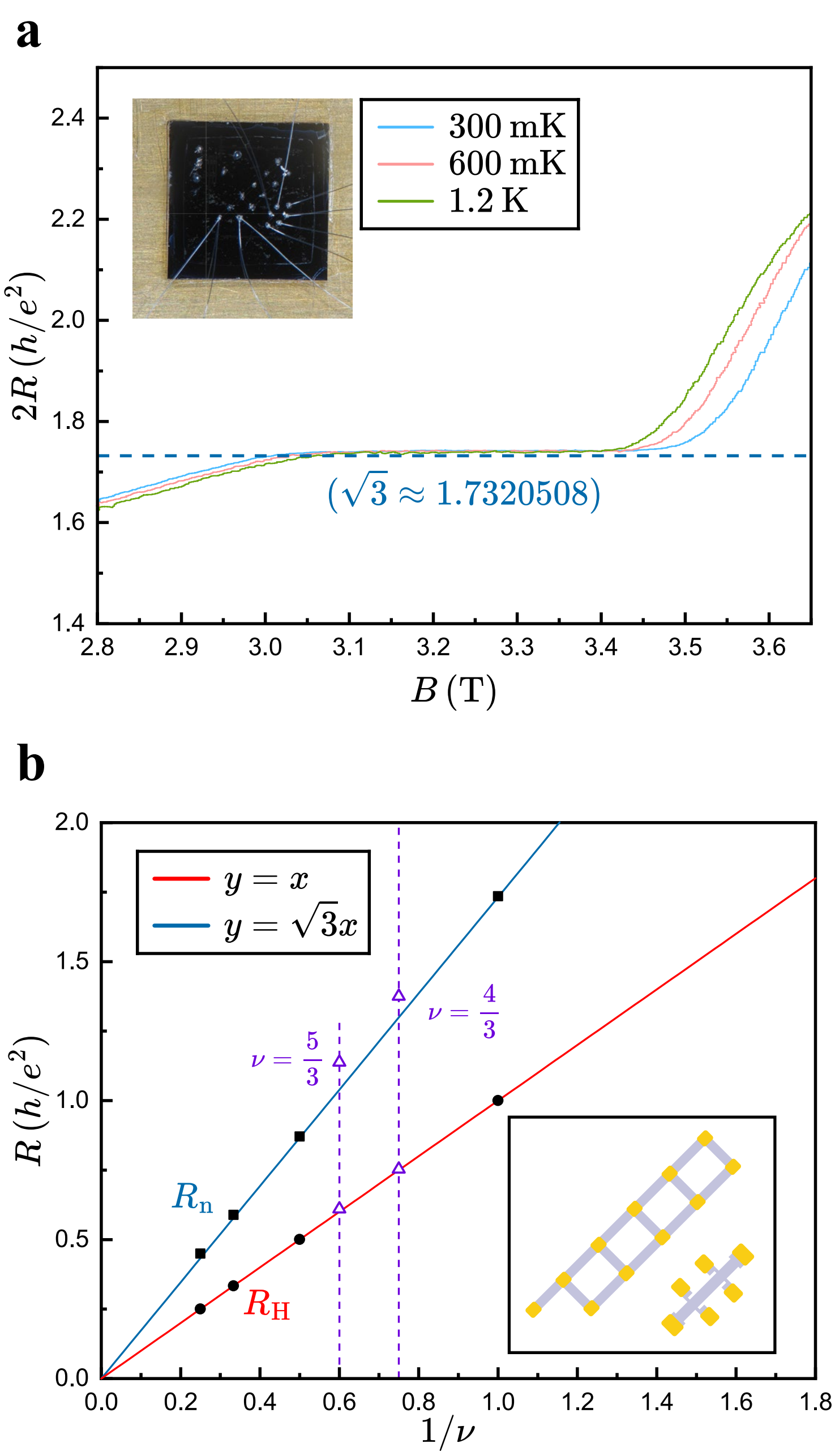

**Figure 4: Experimental observation of irrational quantization. a.** Network resistance as a function of magnetic field measured at 300, 600, and 1200 mK, showing quantized plateaus at values approaching the predicted irrational coefficient $\sqrt{3} \approx 1.7320508$. The inset shows a photo of the fabricated device on a sample holder. **b.** Comparison of the conventional Hall resistance and the network resistance as a function of the inverse filling factor $1/\upsilon$. The data include IQH states at $\upsilon = 1, 2, 3, 4$ and FQH states at $\upsilon = 4/3, 5/3$. The inset shows the device pattern used to realize the quantum network and the control pattern fabricated on the same wafer and the identical procedure.

In this work, we have established a route toward irrational quantization based on the limiting behavior of geometric networks. The central idea is to encode the target value in an infinite geometric structure and to approximate this limit with a finite network whose response converges rapidly as the system size increases. This connects an infinite mathematical construction to an experimentally realizable finite system.

Although different structures and configurations can approach the same limit, their finite-size errors can exhibit distinct behaviors. The asymptotic behavior of the network response near $\boldsymbol{k} = 0$ effectively governs the rate of error decay. Thus, the rate of convergence can be optimized through appropriate structural design. For instance, when the limiting value is rational, a well-designed network can converge to the exact value at a finite size. Our results suggest that increasing the dimensionality of the structure may expand the range of irrational number limits accessible to the network.

For the structure studied here, the normalized resistance approaches $\sqrt{3}$, and the experimental result agrees with the predicted behavior. The small positive deviation is consistent with the contribution of contact resistance. If contact resistance could be eliminated by replacing conventional ohmic contacts with artificial contacts based on density-mismatch junctions[21], more complex networks of edge currents might provide a more direct realization of the theoretical result. More broadly, our approach shows how the limiting properties of geometric networks can provide well-defined quantized responses beyond conventional rational values, offering a framework for exploring quantization associated with non-periodic or infinite structures using finite physical systems. Beyond this specific realization, the framework raises the possibility that quantization can emerge as a well-defined asymptotic limit. A sequence of finite systems may possess rational or otherwise conventional responses at every finite size while converging to an irrational value in the infinite-size limit. As shown in Supplementary Material, Section B and Eq. S15, different irrational quantization values are possible. Whether such asymptotic responses constitute a distinct form of quantization remains an open question. The convergence law may also provide an additional means of characterizing different geometric realizations.

## Acknowledgements

We thank Xincheng Xie, Fuchun Zhang and Xinzheng Li for their discussions. The work at Peking University is supported by the National Key Research and Development Program of China (Grant No. 2021YFA1401900), the National Natural Science Foundation of China (12141001), and the Quantum Science and Technology—National Science and Technology Major Project (Grant No. 2021ZD0302600). The Princeton portion of this work is funded in part by the Gordon and Betty Moore Foundation EPiQS initiative, Grant GBMF14260.

## References

1. Klitzing, K. v., Dorda, G. & Pepper, M. New method for high-accuracy determination of the fine-structure constant based on quantized Hall resistance. *Phys. Rev. Lett.* **45,** 494–497 (1980).
2. Halperin, B. I. Quantized Hall conductance, current-carrying edge states, and the existence of extended states in a two-dimensional disordered potential. *Phys. Rev. B* **25,** 2185–2190

(1982).
3. Thouless, D. J., Kohmoto, M., Nightingale, M. P. & den Nijs, M. Quantized Hall conductance in a two-dimensional periodic potential. *Phys. Rev. Lett.* **49,** 405–408 (1982).
4. Tsui, D. C., Stormer, H. L. & Gossard, A. C. Two-dimensional magnetotransport in the extreme quantum limit. *Phys. Rev. Lett.* **48,** 1559–1562 (1982).
5. Nayak, C., Simon, S. H., Stern, A., Freedman, M. & Das Sarma, S. Non-Abelian anyons and topological quantum computation. *Rev. Mod. Phys.* **80,** 1083–1159 (2008).
6. Willett, R. L. The quantum Hall effect at 5/2 filling factor. *Rep. Prog. Phys.* **76,** 076501(2013).
7. Lin, X., Du, R. R. & Xie, X. C. Recent experimental progress of fractional quantum Hall effect: 5/2 filling state and graphene. *Natl. Sci. Rev.* **1,** 564–579 (2014).
8. Jiang, N. & Wan, X. Recent progress on the non-Abelian ν =5/2 quantum Hall state. *AAPPS Bull.* **29,** 58(2019).
9. Landauer, R. Spatial Variation of Currents and Fields Due to Localized Scatterers in Metallic Conduction. *IBM J Res. Dev.* **1,** 223–231 (1957).
10. Büttiker, M., Imry, Y., Landauer, R. & Pinhas, S. Generalized many-channel conductance formula with application to small rings. *Phys. Rev. B* **31,** 6207–6215 (1985).
11. Wilczek, F. Quantum mechanics of fractional-spin particles. *Phys. Rev. Lett.* **49,** 957–959 (1982).
12. Laughlin, R. B. Anomalous quantum Hall effect: An incompressible quantum fluid with fractionally charged excitations. *Phys. Rev. Lett.* **50,** 1395–1398 (1983).
13. Arovas, D., Schrieffer, J. R. & Wilczek, F. Fractional statistics and the quantum Hall effect. *Phys. Rev. Lett.* **53,** 722–723 (1984).
14. Wen, X.-G. *Quantum Field Theory of Many-Body Systems: From the Origin of Sound to an Origin of Light and Electrons* (Oxford Univ. Press, 2007).
15. Jain, J. K. Composite-fermion approach for the fractional quantum Hall effect. *Phys. Rev. Lett.* **63,** 199–202 (1989).
16. Jain, J. K. *Composite Fermions* (Cambridge Univ. Press, 2007).
17. Chen, Z. et al. Cascade of fractional quantum Hall states in 2D system. *Natl. Sci. Rev.* **13,** nwag079 (2026).
18. Levine, D. & Steinhardt, P. J. Quasicrystals. I. Definition and structure. *Phys. Rev. B* **34,** 596–616 (1986).
19. Hofstadter, D. R. Energy levels and wave functions of Bloch electrons in rational and irrational magnetic fields. *Phys. Rev. B* **14,** 2239–2249 (1976).
20. Cserti, J., Széchenyi, G. & Dávid, G. Uniform tiling with electrical resistors. *J. Phys. A* **44,** 215201 (2011).
21. Yan, J. et al. Anomalous quantized plateaus in two-dimensional electron gas with gate confinement. *Nat. Commun.* **14,** 1758 (2023).

# Supplementary A: Integral calculation in the continuous case

In a periodic network composed of linear elements with a finite number of connections per node, signals can flow through the network, and physical quantities can be read out. The formulation can be generalized to an arbitrary spatial dimension $d$, as indicated by Eq. 1 in the main text, while the specific form of the network operator depends on the component type and frequency. In the specific case of circuit networks, $Y$ is analogous to the Laplacian matrix, and its dimensionality matches the number of nodes within the primitive cell. A concise way to write this relation in real space is the following Eq. S1.

$$Y(\boldsymbol{r}-\boldsymbol{r}')V(\boldsymbol{r}')=I(\boldsymbol{r}) \qquad \text{(S1)}$$

Here, $\boldsymbol{r}'$ needs to be summed at all possible positions and $Y$ serves as the admittance matrix. Due to translational symmetry, $Y$ depends only on the relative displacement $\boldsymbol{r}-\boldsymbol{r}'$. Under Fourier transform, it can be expressed in reciprocal space.

$$Y(\boldsymbol{k})V(\boldsymbol{k})=I(\boldsymbol{k}) \qquad \text{(S2)}$$

Generally, if $Y(\boldsymbol{k})$ is invertible, a current $I(\boldsymbol{r})$ in real space can be Fourier transformed to $I(\boldsymbol{k})$, so that a response $V(\boldsymbol{k})$ is obtained in reciprocal space. Then, the equivalent resistance is obtained by transforming $V(\boldsymbol{k})$ back to real space. When normalizing the primitive cell size parameters, we write $I(\boldsymbol{r})=I\boldsymbol{p}(\boldsymbol{r})$ in Eq. 1.

As an example, we consider the situation in Fig. 1. Assume that each edge is equivalent to a pure resistor $R_0$. The positions of the two external detection nodes in real space are $(a,b)$ and $(c,d)$, respectively.

$$\boldsymbol{p}(\boldsymbol{k})=e^{-i(k_x a+k_y b)}-e^{-i(k_x c+k_y d)} \qquad \text{(S3)}$$

It's originated from the form of real-space $\boldsymbol{p}(\boldsymbol{r})=\delta_{\boldsymbol{r},(a,b)}-\delta_{\boldsymbol{r},(c,d)}$. Positive sign means current flowing in and negative sign means current flowing out. Express $Y$ in its real-space form according to Eq. S1.

$$Y(\boldsymbol{r})=\frac{1}{R_0}\left(4\delta_{\boldsymbol{r},0}-\delta_{\boldsymbol{r},\boldsymbol{e}_x}-\delta_{\boldsymbol{r},-\boldsymbol{e}_x}-\delta_{\boldsymbol{r},\boldsymbol{e}_y}-\delta_{\boldsymbol{r},-\boldsymbol{e}_y}\right) \qquad \text{(S4)}$$

Here $\boldsymbol{e}_x=(1,0)$ and $\boldsymbol{e}_y=(0,1)$. The four connections between adjacent nodes contribute to Eq. S4, including $\boldsymbol{r}\leftrightarrow\boldsymbol{r}+\boldsymbol{e}_x$, $\boldsymbol{r}\leftrightarrow\boldsymbol{r}-\boldsymbol{e}_x$, $\boldsymbol{r}\leftrightarrow\boldsymbol{r}+\boldsymbol{e}_y$, $\boldsymbol{r}\leftrightarrow\boldsymbol{r}-\boldsymbol{e}_y$. Cause there is only one type of node in the network, the matrix has one order. Then applying a Fourier transform, one can obtain

$$Y(\boldsymbol{k})=\frac{1}{R_0}\left(4-2\cos k_x-2\cos k_y\right) \qquad \text{(S5)}$$

We take a special case for the calculation that $(a,b)=(0,0),(c,d)=(1,1)$, so that $\boldsymbol{p}(\boldsymbol{k})=1-e^{-i(k_x+k_y)}$. Given that the read and write nodes are the same, the form of contribution at a special $\boldsymbol{k}$ becomes $\boldsymbol{p}^\dagger(\boldsymbol{k})Y^{-1}(\boldsymbol{k})\boldsymbol{p}(\boldsymbol{k})$ like that in Eq. 1 where the definition is $R_\infty=\frac{\Delta V(r)}{I}=\frac{\boldsymbol{p}^\dagger(r)V(r)}{I}=\int_{\mathrm{BZ}}\frac{\mathrm{d}^d\boldsymbol{k}}{(2\pi)^d}\frac{\boldsymbol{p}^\dagger(\boldsymbol{k})V(\boldsymbol{k})}{I}=\int_{\mathrm{BZ}}\frac{\mathrm{d}^d\boldsymbol{k}}{(2\pi)^d}\boldsymbol{p}^\dagger(\boldsymbol{k})Y^{-1}(\boldsymbol{k})\boldsymbol{p}(\boldsymbol{k})$. Here $\boldsymbol{p}^\dagger(\boldsymbol{k})$ represents Hermitian conjugate of $\boldsymbol{p}(\boldsymbol{k})$. The equivalent resistance is given by Eq. S6.

$$R_\infty = R_0 \int_{-\pi}^{\pi} \int_{-\pi}^{\pi} \frac{\mathrm{d}k_x \mathrm{d}k_y}{(2\pi)^2} \frac{1-\cos(k_x+k_y)}{2-\cos k_x - \cos k_y} = \frac{2}{\pi} R_0 \tag{S6}$$

This calculation method is general and can be applied to other network configurations following the same steps. In finite cases, the integral reduces to a summation, and possible singularities in the denominator similar to Eq. S6 give rise to error terms.

# Supplementary B: Finite-size error analysis

When approximating an infinite-sized network using a finite-sized one, the equivalent resistance is expected to approach a well-defined limiting value as the network size increases. This limiting value is determined by the underlying infinite network, whereas the finite-size error depends on how the finite network approaches the infinite limit. However, the error associated with finite size depends on the manner of convergence. It is foreseeable that boundary conditions will become negligible in the infinite-size limit, so we adopt periodic boundary conditions for convenience. Based on the above constraints, the finite-size version of Eq. 1 is written as

$$F_{p,L}(\boldsymbol{k}) = \frac{1}{L^d} \sum_{\boldsymbol{k} \in K_L} \boldsymbol{p}^\dagger(\boldsymbol{k}) Y^+(\boldsymbol{k}) \boldsymbol{p}(\boldsymbol{k}) \tag{S7}$$

Here $K_L$ means the set of lattice points of a finite-size network's Brillouin zone. $Y(\boldsymbol{k})$ is usually singular at $\boldsymbol{k}=0$ because a global shift in electric potential doesn't alter the current. Moore-Penrose pseudoinverse is necessary at $\boldsymbol{k}=0$ while it remains the normal inverse at $\boldsymbol{k} \neq 0$, which is written as $Y^+(\boldsymbol{k})$. In the infinite case, the fact that the measure of $\boldsymbol{k}=0$ is zero means its contribution doesn't matter, so even if a normal inverse $Y^{-1}(\boldsymbol{k})$ at $\boldsymbol{k}=0$ is singular, Eq. 1 still holds. In the finite-size case, employing a hypercube with side length $L$ facilitates the analysis of error growth ways, similar formulas can be obtained for other shapes as well. By following this approach, we can calculate rigorous finite-size results for the case shown in Fig. 2a. $3R_0$ comes from the practical configuration in Fig. 2a.

$$Y(\boldsymbol{r}) = \frac{1}{3R_0} \begin{pmatrix} 3\delta_{r,0} - \delta_{r,1} - \delta_{r,-1} & -\delta_{r,0} \\ -\delta_{r,0} & 3\delta_{r,0} - \delta_{r,1} - \delta_{r,-1} \end{pmatrix} \tag{S8}$$

Here $Y$'s order is two because two different nodes exist in Fig. 2a. The diagonal terms originate from the connections of the same type, while the off-diagonal terms are governed by the connections between the nodes of different types. Then, $Y(\boldsymbol{k})$ can be obtained as

$$Y(\boldsymbol{k}) = \frac{1}{3R_0} \begin{pmatrix} 3-2\cos k & -1 \\ -1 & 3-2\cos k \end{pmatrix} \tag{S9}$$

From this expression, it can be seen that it is singular at $\boldsymbol{k}=0$, and its pseudoinverse is

$$Y^+(\boldsymbol{k}=0) = \frac{3R_0}{4} \begin{pmatrix} 1 & -1 \\ -1 & 1 \end{pmatrix} \tag{S10}$$

Here $\boldsymbol{p}(\boldsymbol{r}) = (\delta_{r,0}, -\delta_{r,0})^{\mathrm{T}}$ so $\boldsymbol{p}(\boldsymbol{k}) = (1, -1)^{\mathrm{T}}$ (T means transpose, and the choice of zero point eliminates the complex phase in Eq. S3), so $\boldsymbol{p}^{\dagger}(\boldsymbol{k})Y^{+}(\boldsymbol{k})\boldsymbol{p}(\boldsymbol{k}) = \frac{3R_0}{2-\cos k}$. Despite using a pseudoinverse, $\boldsymbol{p}^{\dagger}(\boldsymbol{k})Y^{+}(\boldsymbol{k})\boldsymbol{p}(\boldsymbol{k})$ at $\boldsymbol{k} = 0$ is exactly equal to the value $\boldsymbol{p}^{\dagger}(\boldsymbol{k})Y^{-1}(\boldsymbol{k})\boldsymbol{p}(\boldsymbol{k})$ obtained by extending the expression calculated for the case where $\boldsymbol{k} \neq 0$. This causes an exponential decay of the error term. We use $k = \frac{2\pi m}{L}$ to write the equation.

$$R_L = \frac{3R_0}{L}\sum_{m=0}^{L-1}\frac{1}{2-\cos\left(\frac{2\pi m}{L}\right)} = \sqrt{3}R_0\left[1 + \frac{2}{\left(2+\sqrt{3}\right)^L - 1}\right] \tag{S11}$$

Similar calculations can also be performed on Fig. 2b. The difference is that the singularity of this network near $\boldsymbol{k} = 0$ introduces a power-law decay term not present in the previous case.

$$R_L = \frac{3R_0}{L}\sum_{m=0}^{L-1}\frac{1}{2+\cos\left(\frac{2\pi m}{L}\right)} - \frac{R_0}{L} = \sqrt{3}R_0\left[1 + \frac{2}{\left(-2-\sqrt{3}\right)^L - 1}\right] - \frac{R_0}{L} \tag{S12}$$

Here $\boldsymbol{p}(\boldsymbol{r}) = \delta_{r,0} - \delta_{r,1}, \boldsymbol{p}(\boldsymbol{k}) = 1 - e^{-ik}$ and $Y(\boldsymbol{k}) = \frac{1}{3R_0}[3 - 2\cos k - \cos(2k)]$ (there is only one node type). The difference lies in the fact that $\boldsymbol{p}^{\dagger}(\boldsymbol{k})Y^{+}(\boldsymbol{k})\boldsymbol{p}(\boldsymbol{k}) = 0$ at $\boldsymbol{k} = 0$, which differs the result by extending the $\boldsymbol{k} \neq 0$ expression, inducing a single count of $\sim \frac{1}{L^d}$. Consequently, a power-law decay appears.

Power-law decay can frequently emerge even in networks that admit exponentially decaying solutions. We conjecture that the latter may be a universal phenomenon. For example, if the measurement is performed between two adjacent nodes on the same outer ring in Fig. 2a, the general form of $R_L$ becomes Eq. S13.

$$R_L = 3R_0\left(1 - \frac{1}{2L}\right) - \frac{3R_0}{2L}\sum_{m=0}^{L-1}\frac{1}{2-\cos\left(\frac{2\pi m}{L}\right)} = \left[3 - \frac{\sqrt{3}}{2} - \frac{3}{2L} - \frac{\sqrt{3}}{\left(2+\sqrt{3}\right)^L - 1}\right]R_0 \tag{S13}$$

It is similar to follow the approach. Here $\boldsymbol{p}(\boldsymbol{k}) = (\delta_{r,0} - \delta_{r,1}, 0)^{\mathrm{T}}, \boldsymbol{p}(\boldsymbol{k}) = (1 - e^{-ik}, 0)^{\mathrm{T}}$ while $Y(\boldsymbol{k})$ stays the same in Eq. S9. The reason for the emergence of a power-law decay here is similar to that in Eq. S12.

To further obtain a compact network capable of approaching $\sqrt{3}R_0$, we examine the pattern shown in Fig. 3a. The convergence is even faster than that described by Eq. S11.

$$R_L = \sqrt{3}R_0\left[1 + \frac{2}{\left(2+\sqrt{3}\right)^{2(L+1)} - 1}\right] \tag{S14}$$

The pattern shown in the Fig. 3a is obtained through a halving adjustment of Fig. 2a. The error calculation method for the irrational number part can be equivalent to Eq. S11. This can also be obtained using the recursive analysis shown in Fig. 3a. We can examine the convergence behavior as $L$ varies for Fig. 3a, as summarized in Table S1.

**Table S1: The convergence rate of the pattern in Fig. 3a.**

| $L$ | 1 | 2 | 3 | 4 | 5 |
|---|---|---|---|---|---|
| $E_L/R_\infty$ | $1.0\times10^{-2}$ | $7.4\times10^{-4}$ | $5.3\times10^{-5}$ | $3.8\times10^{-6}$ | $2.7\times10^{-7}$ |

Beyond these cases whose limiting values are composed of rational numbers and an irrational number $\sqrt{3}$, we provide an example with an irrational number that is not $\sqrt{3}$. Just let $R_a=\frac{R_0}{a}$ and $R_b=\frac{R_0}{b}$ in Fig. 2a where $a$ and $b$ both positive rational numbers, we get Eq. S15.

$$R_L=\frac{R_0}{\sqrt{b(b+2a)}}\left[1+\frac{2}{\left(1+\frac{b}{a}+\sqrt{\frac{b}{a}\left(2+\frac{b}{a}\right)}\right)^L-1}\right] \tag{S15}$$

Here $R_\infty=\frac{R_0}{\sqrt{b(b+2a)}}$, so irrational forms involving the square root of any rational number can be obtained by adjusting the parameters. For example, $a=\frac{1}{4}, b=\frac{1}{2}$ makes $R_\infty=\sqrt{2}R_0$, and $a=b=\frac{1}{3}$ restores the result of Fig. 2a.

## Supplementary C: Measurements at fractional fillings

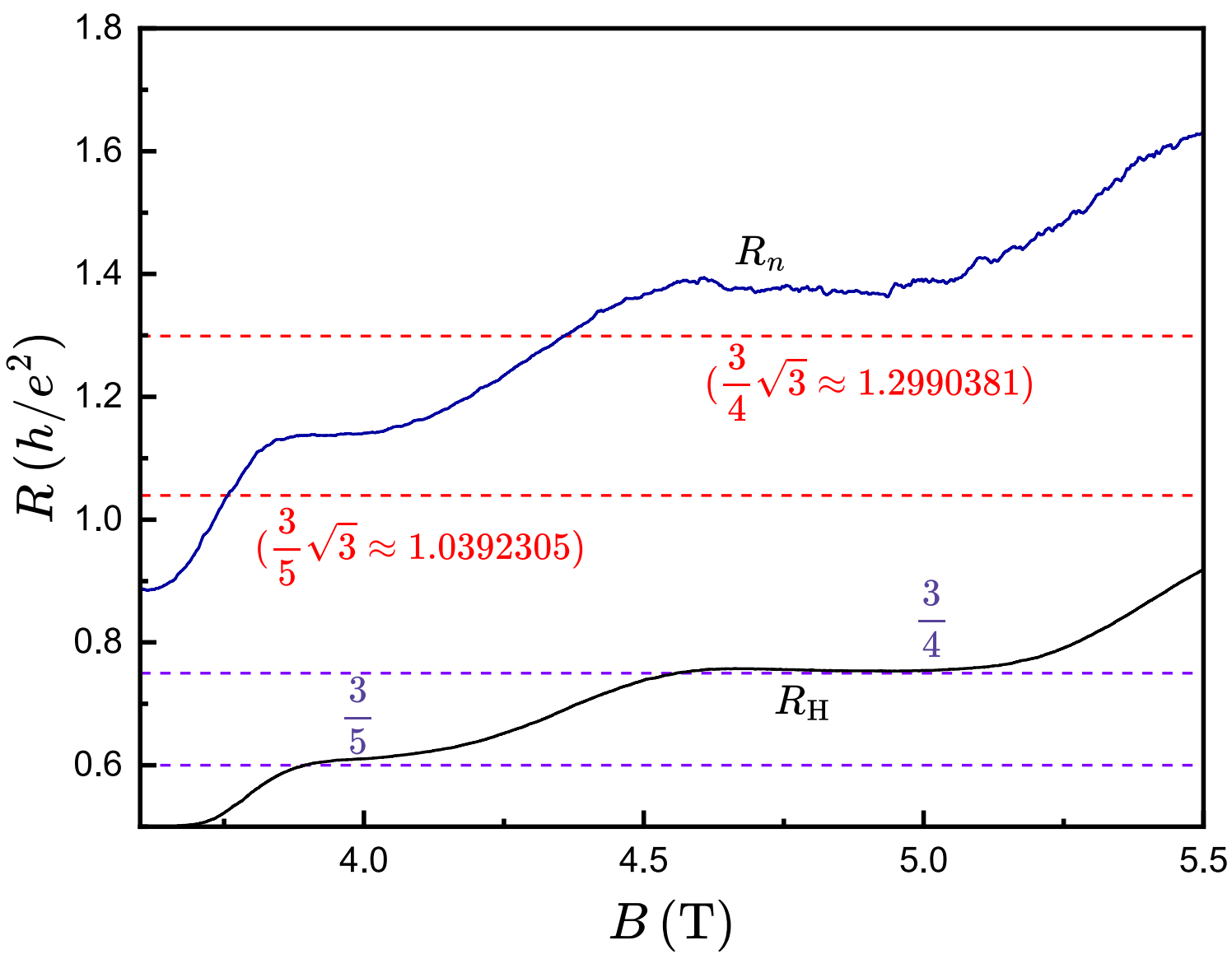

**Figure S1: Conventional Hall resistance and the network resistance measured at fractional filings at around 60 mK.** The conventional Hall resistances are relatively close to the theoretical ones.